\documentclass[10pt,letterpaper]{article}
\usepackage{opex3}
\usepackage{amsmath}
\usepackage{subfigure}

\begin{document}

\title{Incoherently pumped continuous wave optical parametric oscillator broadened by non-collinear phasematching}


\author{Jelle Storteboom$^{1}$, Chris J. Lee$^{1,2}$, Ab F. Nieuwenhuis$^{1}$, Ian D. Lindsay$^{3}$ and Klaus-J. Boller$^{1*}$}

\address{
$^{1}$ Laser Physics and Nonlinear Optics Group, MESA+ Research Institute for Nanotechnology, University of Twente, PO Box 217, 7500AE Enschede, Netherlands\\
$^{2}$FOM Institute for Plasma Physics Rijnhuizen, PO Box 1207, NL-3430 BE  Nieuwegein, the Netherlands (work address: ASML, De Run 6501, 5504 DM Veldhoven)\\
$^{3}$HH Wills Physics Laboratory, School of Physics, Tyndall Avenue, Bristol, BS8 1TL, the United Kingdom\\
$^*$Corresponding author: k.j.boller@tnw.utwente.nl
}	

\begin{abstract}In this paper, we report on a singly resonant optical parametric oscillator (OPO) pumped by an amplified spontaneous emission (ASE) source. The pump focusing conditions allow non-collinear phasematching, which resulted in a 230~nm (190~cm$^{-1}$) spectral bandwidth. Calculations indicate that such phasematching schemes may be used to further broaden OPO spectral bandwidths.\end{abstract}


\section{Introduction}
\noindent The development of broadband, spatially bright continuous wave optical parametric oscillators (OPOs) are of considerable interest. In particular, broadband, temporally incoherent, but spatially bright mid infrared sources are of interest for spectroscopy~\cite{Yabuzaki:1991vk}, speckle free imaging~\cite{Powers:2000ie},  improvements in measurements based on Fourier transform infrared interferometers (FTIR)~\cite{Tillman:2004gx,Adler:2010tq}, optical coherence tomography~\cite{Colley:2007dn}, or for investigations of incoherent optical solitons~\cite{Picozzi:2004ca,Picozzi:2001hp}. In the case of a train of broadband femtosecond pulses, down conversion through difference frequency generation in a short nonlinear optical crystal allows for the generation of signal and idler fields that are spectrally broad. However, the output still has a high degree of temporal coherence---in the sense that the output consists of a regular train of pulses, which maintain a slowly varying phase relationship to one another so that neighboring pulses can exhibit near-complete destructive interference when temporally overlapped---due to the comb-like spectrum consisting of narrowband modes. An alternative is to use continuous wave sources. However, in general this requires longer nonlinear crystals, which, in turn narrows the phasematching bandwidth and precludes the generation of spectrally broad difference frequencies.

Recently Das \emph{et al.}~\cite{Das:2009wza} demonstrated that, if the phasematching conditions are chosen such that the frequency mixing is centered around the location in the tuning curve where $d\lambda_{s}/d\lambda_{p}\approx 0$ ($\lambda_{p}$, $\lambda_{s}$, and $\lambda_{i}$ are the pump, signal and idler wavelengths), then a broadband pump can drive an OPO that has a highly coherent signal (spectrally narrow) and a broadband, temporally incoherent idler~\cite{Das:2009wza}. The reported OPO had an idler spectral bandwidth of 75~nm (64~cm$^{-1}$), centered near 3450~nm, agreeing with their bandwidth calculations.

The OPO demonstrated in~\cite{Das:2009wza} had an optical cavity with a rather large beam waist (76~$\mu$m), which in combination with a weakly focused pump ensured that collinear interactions dominated the OPO's spectral characteristics. The combination of collinear phasematching and $d\lambda_{s}/d\lambda_{p}\approx 0$ cannot, however, be fulfilled for all spectral regions of interest. A more general approach is to use a high-power continuous wave pump and non-collinear phasematching to obtain a spectrally broad idler and a spectrally narrow signal. Here, we demonstrate the simplest case, where a strongly focused pump, in combination with an OPO cavity with a tight signal waist, allows for non-collinear phasematching processes to significantly broaden the output spectrum of the OPO, operating in the region where $d\lambda_{s}/d\lambda_{p}\approx0$. We show that, with a similar input pump spectral bandwidth to that used in~\cite{Das:2009wza}, we obtain an idler spectral bandwidth up to 230~nm (190~cm$^{-1}$).

\section{Experimental setup}
The OPO is pumped by a home-built amplified spontaneous emission (ASE) source that consists of a matched pair of Ytterbium-doped double clad polarization maintaining fibers, similar to the master-oscillator power amplifier systems that we have described elsewhere~\cite{Lindsay:2006tna, Adhimoolam:2006vv}. Using two fibers with identical physical properties (Nufern PLMA-YDF-20/400 6.5~m length, 20~$\mu$m core diameter) for both seeding and amplifying ensures that the seed spectrum optimally matches the amplifier spectrum. Both active fibers were pumped at 980~nm by diode lasers capable of emitting up to 50~W of output power. One fiber was operated at low power (approximately 70~mW) and its ASE output was used to seed the second, which acted as a high power amplifier. A 60 dB optical isolator was placed between the two to prevent feedback from destabilizing the ASE output. In addition, all fiber ends were angle polished to reduce feedback that would destabilize the ASE output of the seed or amplifier. Both were coiled tightly (radius $\sim$6~cm) to increase the losses for higher spatial modes and improve the spatial beam quality of the amplifier output~\cite{Lindsay:2006tna}. The polarized output of the amplifier was then passed through another 60~dB optical isolator and focused into the OPO's nonlinear crystal using an $f$=150~mm focal length lens. The waist in the center of the crystal was calculated to have a 23~$\mu$m diameter (FWHM), which is poorly matched to the waist of the OPO's resonant signal that was calculated to have a diameter of 64~$\mu$m (FWHM), but provided a large spread of wavevectors to maximize noncollinear interactions. 

The OPO is similar with earlier work, where it was used to generate narrowband mid-infrared tunable light~\cite{Lindsay:2006tna}. Briefly, the OPO is based on a 50~mm long periodically poled MgO doped lithium niobate crystal (HC Photonics) with a 0.5$\times$10~mm aperture that is anti-reflection coated for the pump, signal, and idler waves. The crystal was mounted on a brass oven and maintained at a temperature of 77$^{\circ}$C.The crystal has multiple poling periods, however, only the 30~$\mu$m poling period was used for this work. The bow-tie ring resonator (optical length: 260~mm, free spectral range: 940~MHz) consisted of two curved mirrors (radius of curvature of 50~mm) and two flat mirrors, positioned so that the main cavity waist coincided with the center of the crystal and had a confocal parameter of 7.4~mm. All mirrors have a reflectivity in excess of 99.9\% between 1.45 and 1.6~$\mu$m, and are anti-reflection coated for the pump (1--1.1~$\mu$m) and idler (2.7--5~$\mu$m) wavelengths. The bow-tie ring configuration ensures that the pump and idler roundtrip losses are high enough to prevent doubly or triply resonant operation.  

The idler and depleted pump radiation were coupled out through the CaF$_{2}$ output coupler. The pump was then separated from the idler using a dichroic mirror and the idler was collimated using a $f$=150~mm lens for propagation to diagnostic instruments. The idler power was measured using a power meter (Newport power/energy meter model 841-PE with a 818P-030-19 thermopile sensor). The beam was imaged using a mid-infrared camera (Electrophysics PV320), and its spectrum measured using a grating monochromator with a resolution of 1~nm. The monochromator was also used as a spectrometer with the output slit replaced with the mid-infrared camera, which reduced the resolution to 15~nm, but allowed for online recording of the entire idler spectrum. The systematic change to the spectra introduced by the mismatch between the image plane of the monochromator and the cameraÕs sensor was smaller than the resolution of the spectrometer. The idler spectrum as a function of location within the idler beam profile was obtained by placing a pinhole in the beam path between the OPO output and spectrometer input slit. To maintain a fixed propagation direction of the light into the spectrometer input, the location of the pinhole remained fixed and the spatial location of the beam incident to the pinhole was altered by translating a beam steering mirror mounted on a translation stage.

\section{Experimental results}
\begin{figure}[htb]
\centerline{
\subfigure[]{\includegraphics[width=6.5cm]{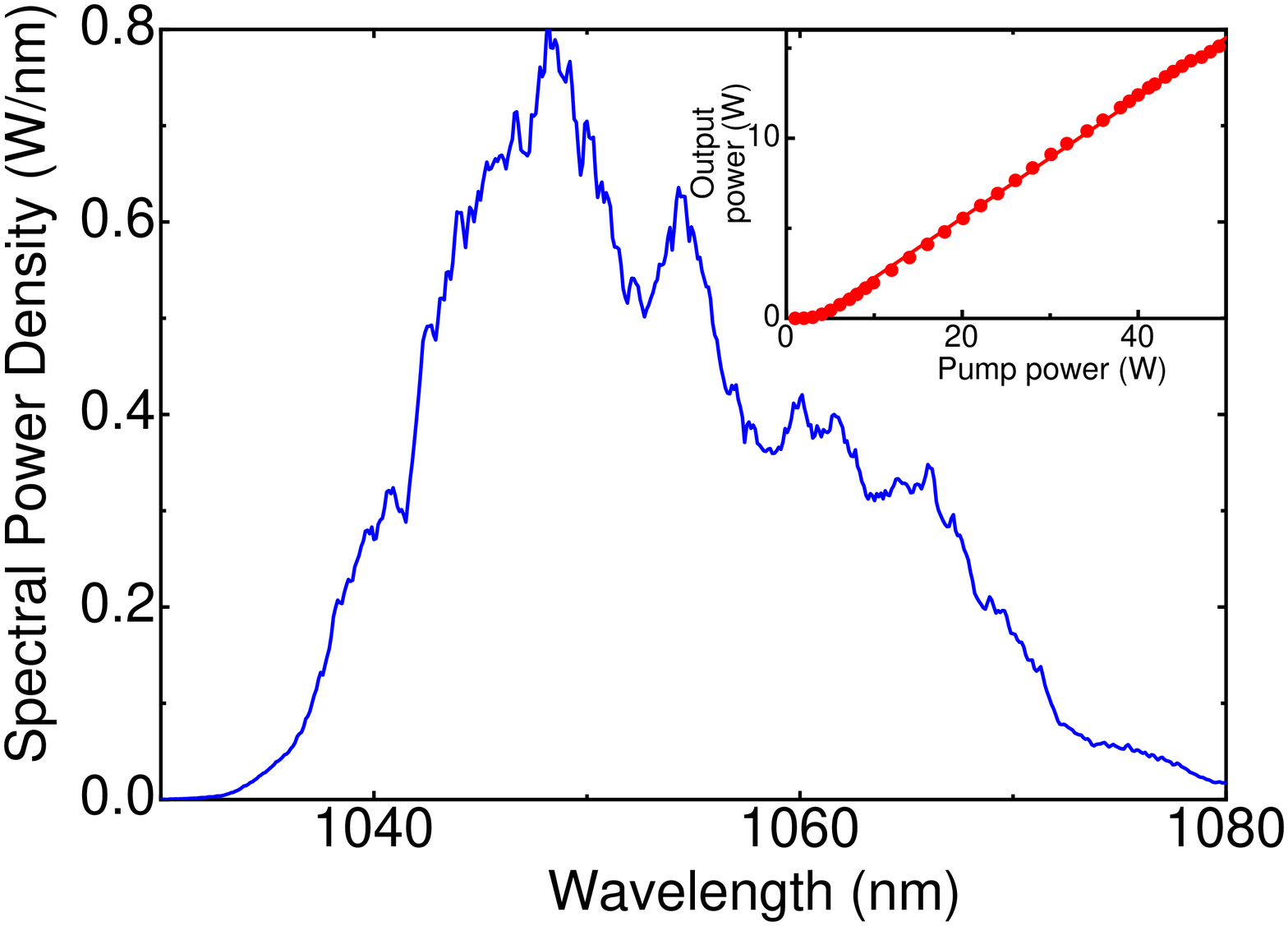}}
\subfigure[]{\includegraphics[width=6.2cm]{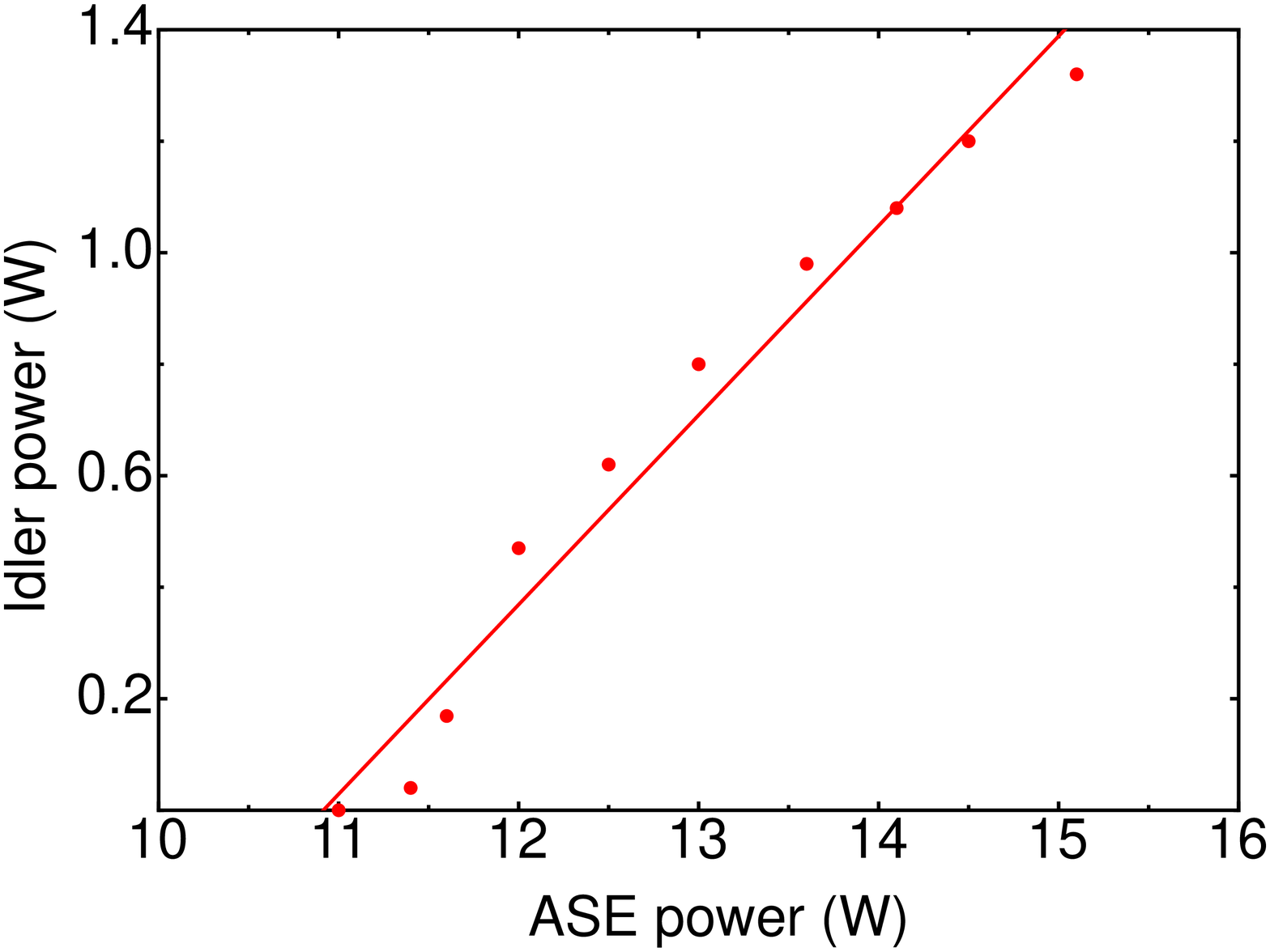}}
}
\caption{Output spectrum of the ASE source (a). Inset of (a) is the ASE output power as a function of the pump diode's optical output power. Idler output power as a function of ASE pump power (b).}
\label{fig:ASE}
\end{figure}

The ASE source was capable of emitting up to 15.3~W of power of which 12.3~W was delivered to the OPO. Fig.~\ref{fig:ASE}(a) shows the spectrum of the ASE source for an output power of 12~W. The spectrum is centered on 1048~nm, has a fullwidth half maximum (FWHM) bandwidth of 15.2~nm, from which we estimate, based on a rectangular spectrum, that the coherence time is 240~fs. The inset of Fig.~\ref{fig:ASE}(a) shows the optical-to-optical power conversion efficiency of the ASE amplifier. The slope efficiency of the ASE amplifier was found to be 33\% while the peak efficiency was 31\%, which is comparable to other fiber amplifier systems~\cite{Adhimoolam:2006vv}.

\begin{figure}[htb]
\centerline{\includegraphics[width=7.5cm]{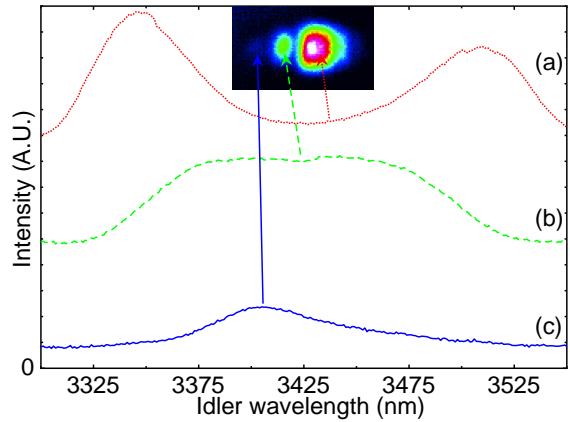}}
\caption{Idler spectrum for different locations in the idler beam profile (excerpted from Media~1). The arrows indicate which lobe of the idler beam the spectrum was obtained from. The spectra (a) and (b) are offset for clarity. The variation across the beam profile from a single-lobed (c) to double-lobed (a) spectrum is explained in the discussion of section~\ref{discussion}.}
\label{fig:OPOSpectrum}
\end{figure}

Fig.~\ref{fig:ASE}(b) shows the measured idler output power of the OPO. We found that the OPO had a threshold of 11~W, a slope efficiency of 33\% and a pump-to-idler conversion efficiency of 8.3\%. The relatively high threshold and low maximum conversion efficiency are due to the poor mode-matching. The advantage of the mode-mismatch is the increased phasematching bandwidth as evidenced by the spectral data and beam profile presented in Fig.~\ref{fig:OPOSpectrum}. The inset shows the idler beam profile, whose multi-lobed nature we will explain below. The signal spectral bandwidth was observed to be less than the 10~pm resolution of the optical spectrum analyzer (ANDO AQ6317). In addition, the spatial profile of the signal mode was qualitatively observed to be well collimated, round and symmetric, as expected for the low-loss signal wave oscillating on the fundamental spatial mode of the OPO optical cavity. 

The idler spectra shown in Fig.~\ref{fig:OPOSpectrum}(a)--(c) are from each lobe of the idler beam, as indicated by the arrows. The lobes appear at a fixed angle with respect to the axis of the pump beam and are oriented along the 0.5~mm direction ($z$-axis) of the crystal. The FWHM spectral bandwidth of the idler without any spatial filtering is 230~nm (190~cm$^{-1}$), obtained for an output power of 1~W (the idler beam profile was recorded at approximately 500~mW, but was found to not vary noticeably for idler powers greater than 300~mW). Although the spectral bandwidth of the idler appears to be larger than that of the pump, about 25\% of the pump power lies outside of the FWHM bandwidth, yet still contributes to the frequency conversion process, accounting for the difference. 

Media 1 illustrates the spatial dependence of the idler spectrum in the three-lobed beam. The movie shows the variation of the idler spectrum recorded with the camera-based spectrometer while the idler beam was translated across the pinhole.

\section{Discussion}
\label{discussion}
To understand the idler spectrum, we must take into consideration that the pump is strongly focused, while the signal is bounded by the cavity, meaning that both collinear and non-collinear phasematching are taken into account. In the following, the idler output spectrum is modeled, based on phasematching considerations, the pump input spectrum, and the pump divergence.  A full model of the OPO is expected to be rather complex as it would have to include, e.g., the influence of pump depletion. Here we extend the simplified model described in ref~\cite{Das:2009wza} to show that the observed spectral properties can largely be addressed to phase matching non-collinear phasematching involving the pump divergence in addition to the pump input spectrum. We approximate the signal with a plane wave, propagating along the axis of the crystal. The pump and idler are a summation of plane waves, with each plane wave component traveling at angles, $\theta$ (pump) and $\phi$ with respect to the axis of the crystal. The phasematching condition can then be broken into longitudinal (subscript $z$) and transverse (subscript $r$) components
\begin{align}
	\label{eq:transverse}
	dk_{r} &= k_{p}\sin\theta - k_{i}\sin\phi\\
	\label{eq:longitudinal}
	dk_{z} &= k_{p}\cos\theta - k_{s} - k_{i}\cos\phi - 1/\Lambda
\end{align}
where $k_{p,s,i}$ are the pump, signal, and idler wavevectors and $\Lambda$ is the crystal poling period.

The divergence half angle of the pump, which closely approximated a Gaussian, was measured to be 50~mrad, which reduces to 8~mrad inside the PPLN crystal, providing the range of values over which $\theta$ varies. For a given pump wavelength and plane wave component, equations~\ref{eq:transverse} and \ref{eq:longitudinal} determine the idler plane wave component and angle, $\phi$ that minimizes the total phase-mismatch, $\Delta k = \sqrt{dk_{z}^{2} + dk_{r}^{2}}$. 

By considering the intensity as a function of radius at the pump input focusing lens, and calculating the propagation angles using a ray-optics approximation, the intensity of each plane wave component of the pump mode, as a function of $\theta$, is given by
\begin{equation}
	I_{p}(\theta,\omega) = I_{0}(\omega_{p})\exp\big(\frac{-f^{2}\tan^{2}\theta}{r_{l}^{2}}\big)
	\label{eq:intensityVsAngle}
\end{equation}
where $I_{0}(\omega_{p})$ is the ASE emission intensity at frequency $\omega$. $f$ is the focal length of the pump focusing lens and $r_{l}$ is the 1/$e^{2}$ radius of the pump beam at the focusing lens. The OPO gain as a function of the divergences and pump frequency is given by:

\begin{equation}
	g(\theta, \phi, k_{p}) = \bigg(I_{p}(\theta,\omega)\frac{\sin(\frac{\Delta kl)}{2}}{\frac{\Delta kl}{2}}\bigg)^{2}
	\label{eq:singleFreqGain}
\end{equation}

The total gain spectrum as a function of the pump input spectrum is then obtained by integrating over $\theta$.
\begin{align}
	G( \omega_{p}, \phi) &= I_{0}(\omega_{p})\int_{\theta=0}^{\theta=\theta_{max}}g(\theta,\phi,k_{p})d\theta
	\label{eq:gainSpectrum}
\end{align}
where $\theta_{max}$ is the maximum half-angle divergence of the pump. Note, however, that $\theta$ and $\phi$ are not independent of each other, as indicated by equations~\ref{eq:transverse} and \ref{eq:longitudinal}. Effectively, one must input $\theta$ into equations~\ref{eq:transverse} and \ref{eq:longitudinal} and obtain a value for $\phi$ that minimizes $\Delta k$.
\begin{figure}
	\centerline{
	 \subfigure[]{\includegraphics[width=6cm]{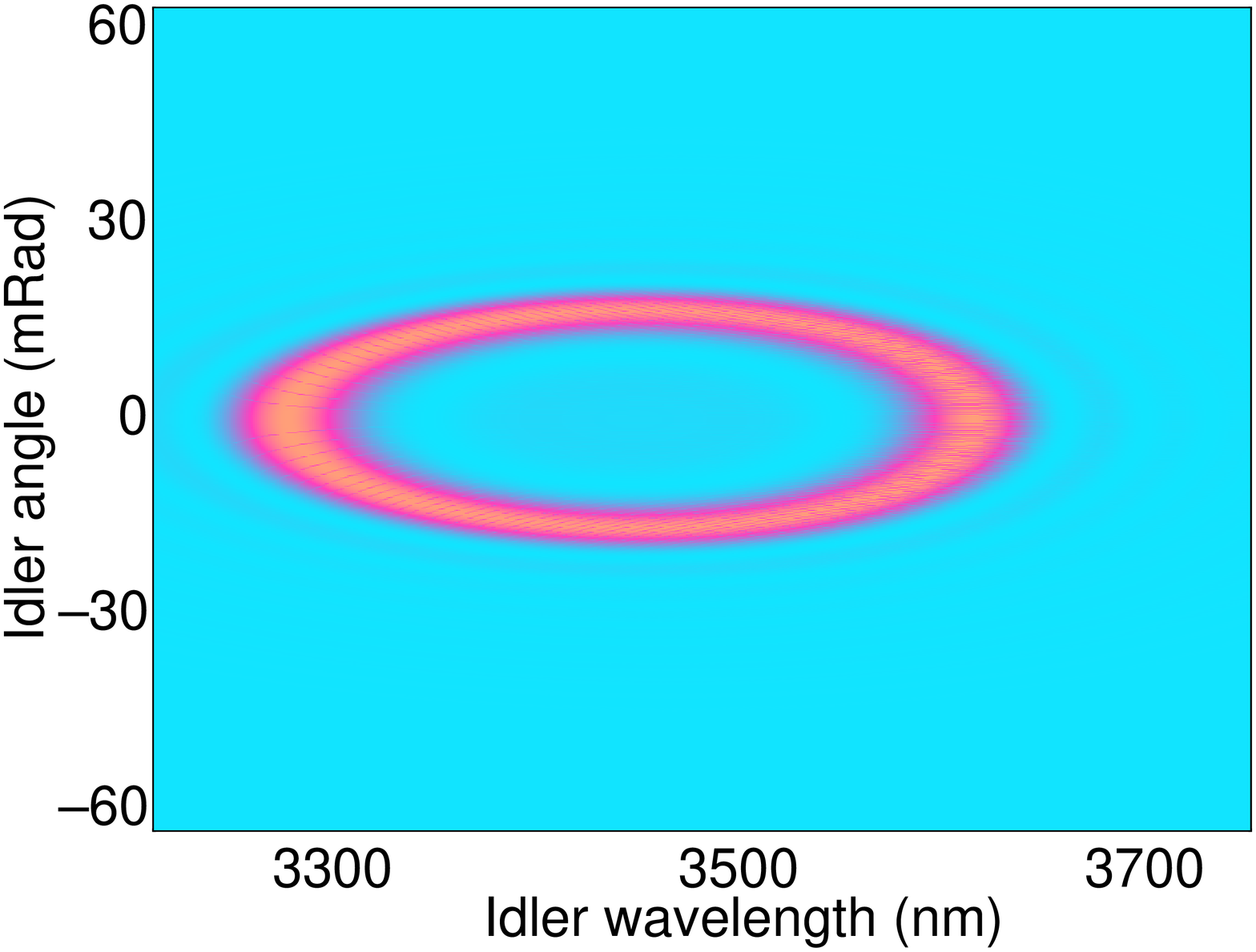}}
	 \subfigure[]{\includegraphics[width=6.5cm]{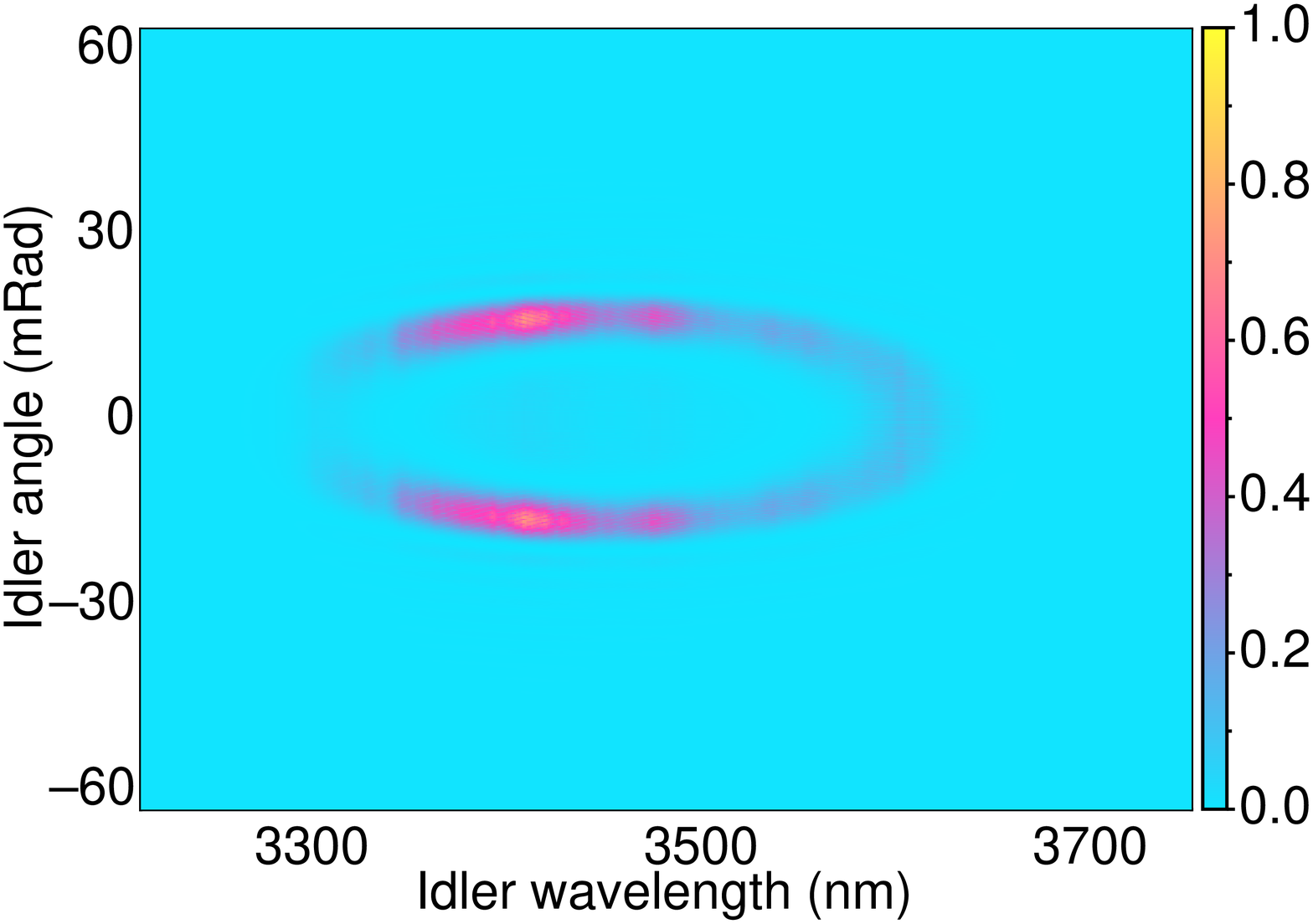}}\\
	 }
	 \caption{Calculated OPO idler spectra as a function of idler wavelength and emission angle. (a) is the wavelength range for which phasematching is possible, taking into account the pump divergence. Subfigure (b) shows a reduced gain region when the pump spectrum from Fig.~\ref{fig:ASE} is taken into account in via equation~\ref{eq:gainSpectrum}. 
	 }
	 \label{fig:phaseMatchingBandwidth}
\end{figure}

Fig.~\ref{fig:phaseMatchingBandwidth}(a) shows the ideal phasematching spectrum of the OPO as a function of $\phi$ and idler wavelength (i.e., the pump is assumed to have a rectangular emission spectrum \emph{and} a uniform intensity contribution to over the range $\phi$=0 to $\phi$ = $\phi_{max}$). One can see that the ideal phasematching bandwidth is very broad, extending from 3270 to 3650~nm. Once the Gaussian spatial intensity distribution of the pump mode and the pump spectrum (Fig.~\ref{fig:ASE}(a)) are taken into account, the OPO gain bandwidth is reduced to between 120 and 150~nm, as shown in Fig.~\ref{fig:phaseMatchingBandwidth}(b). It can also be seen that the idler spectrum is asymmetric, which is due to the asymmetry of the pump spectrum.

The calculations, however, do not explain the idler beam profile's asymmetry. There are several possible sources of asymmetry, such as, incomplete periodic poling, a temperature gradient across the crystal, a tilt of the crystal with respect to the optical axis, or a spatial variation in the pump spectrum. But, while these could be ruled out, we note that the divergence angles of the idler are large enough that the idler can be expected to reflect off the sides of the crystal. The calculated idler beam diameter (FWHM) at the exit of the PPLN crystal is 0.8~mm. The distance from the idler waist to where it has FWHM equaling the width of the PPLN crystal width is 16~mm. In our experimental setup, some of the idler would be incident on a PPLN/glass interface, while some would be incident on a PPLN/rough brass interface. It is likely that the PPLN/brass interface has additional scattering losses. These losses are not expected to be strongly wavelength dependent, but provide a loss mechanism that depends on the divergence angle of the idler. Since the idler spectrum depends on the divergence, the losses at the interface introduce some spectral selectivity. The scattering losses deplete the part of the idler that has the largest negative emission angle (in the coordinates of Fig.~\ref{fig:phaseMatchingBandwidth}), introducing an asymmetry into the idler beam profile.

\begin{figure}
	\centerline{
	 \subfigure[]{\includegraphics[width=6cm]{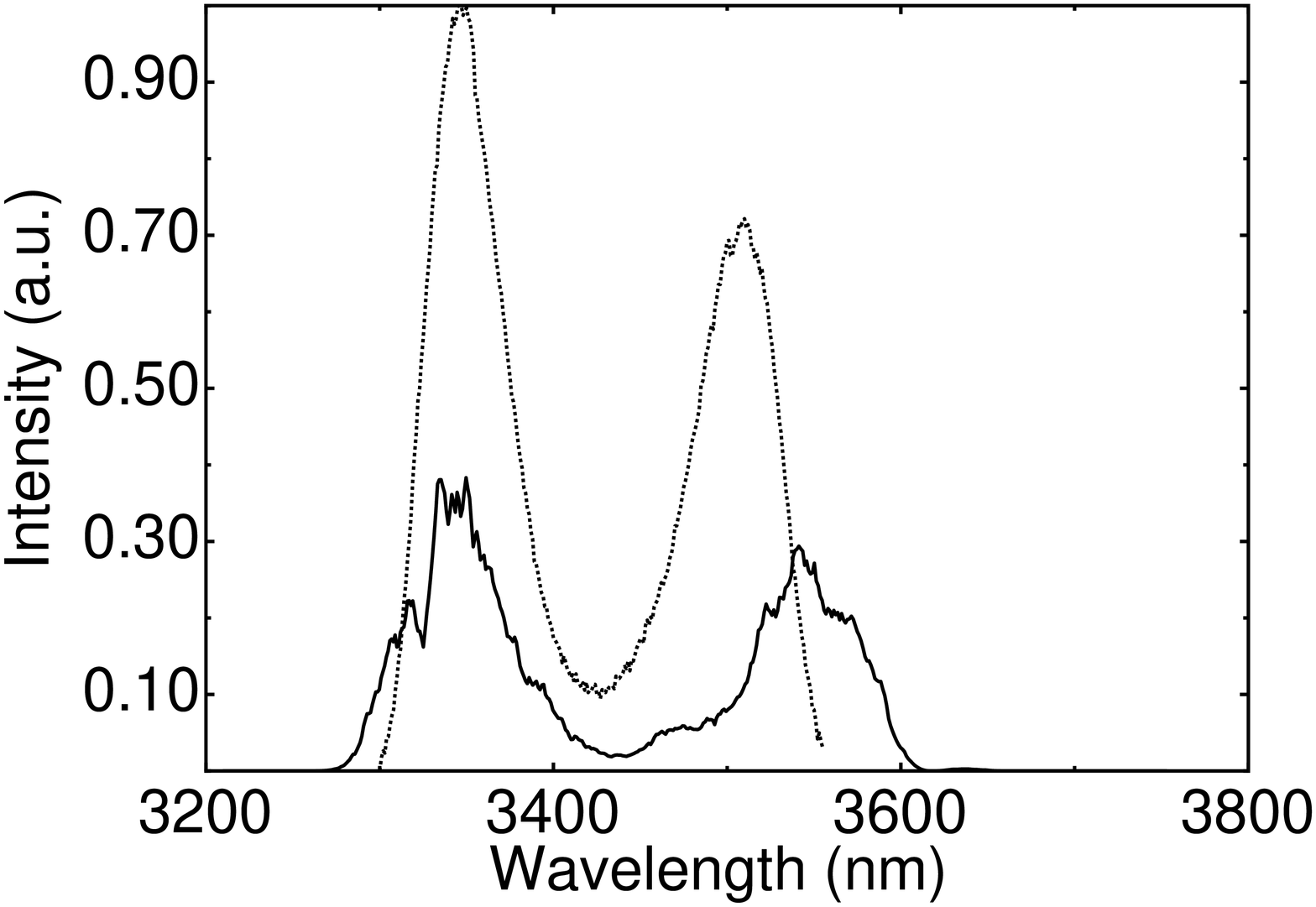}}
	 \subfigure[]{\includegraphics[width=5.8cm]{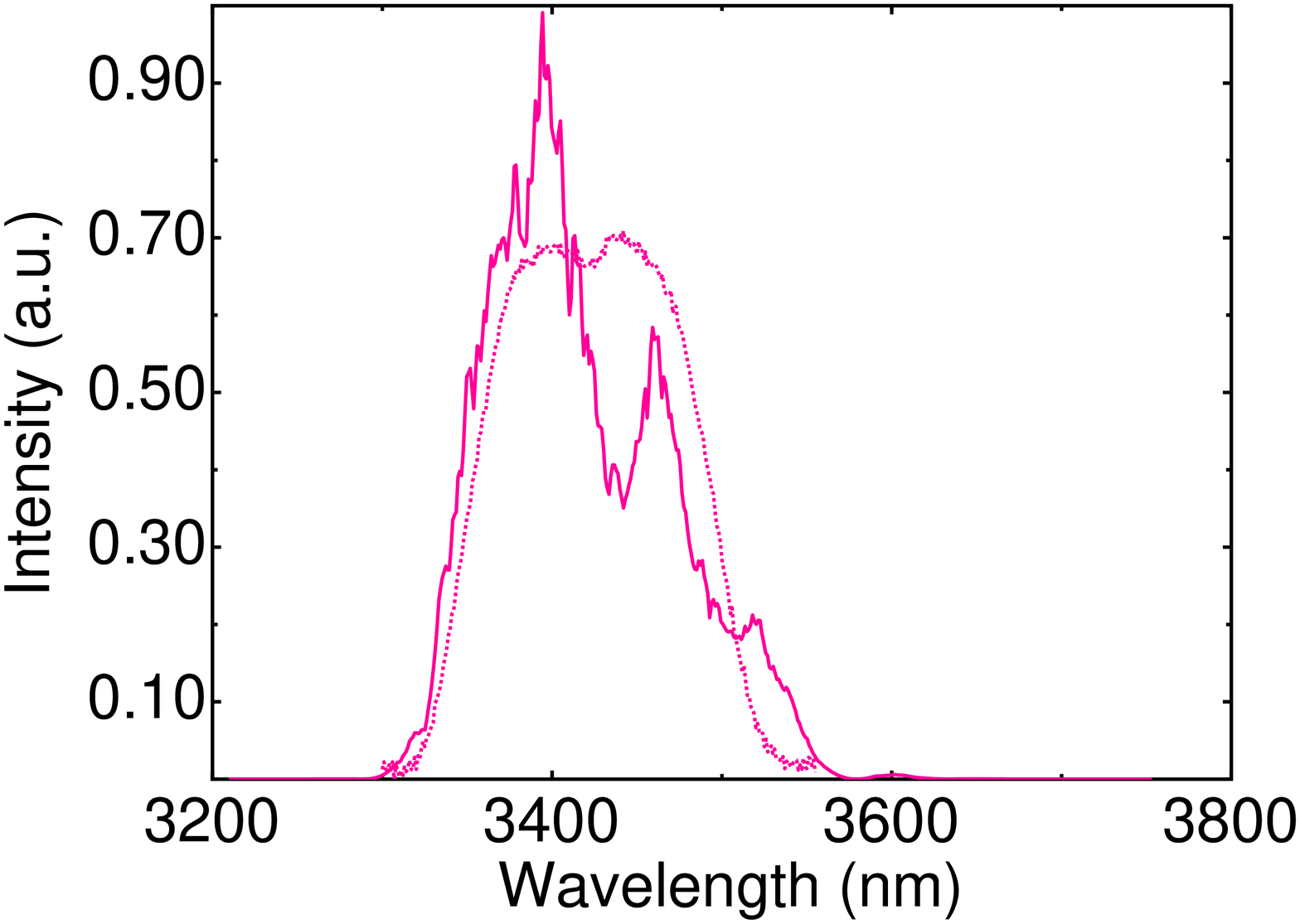}}\\
	 }
	 \centerline{
	 \subfigure[]{\includegraphics[width=6cm]{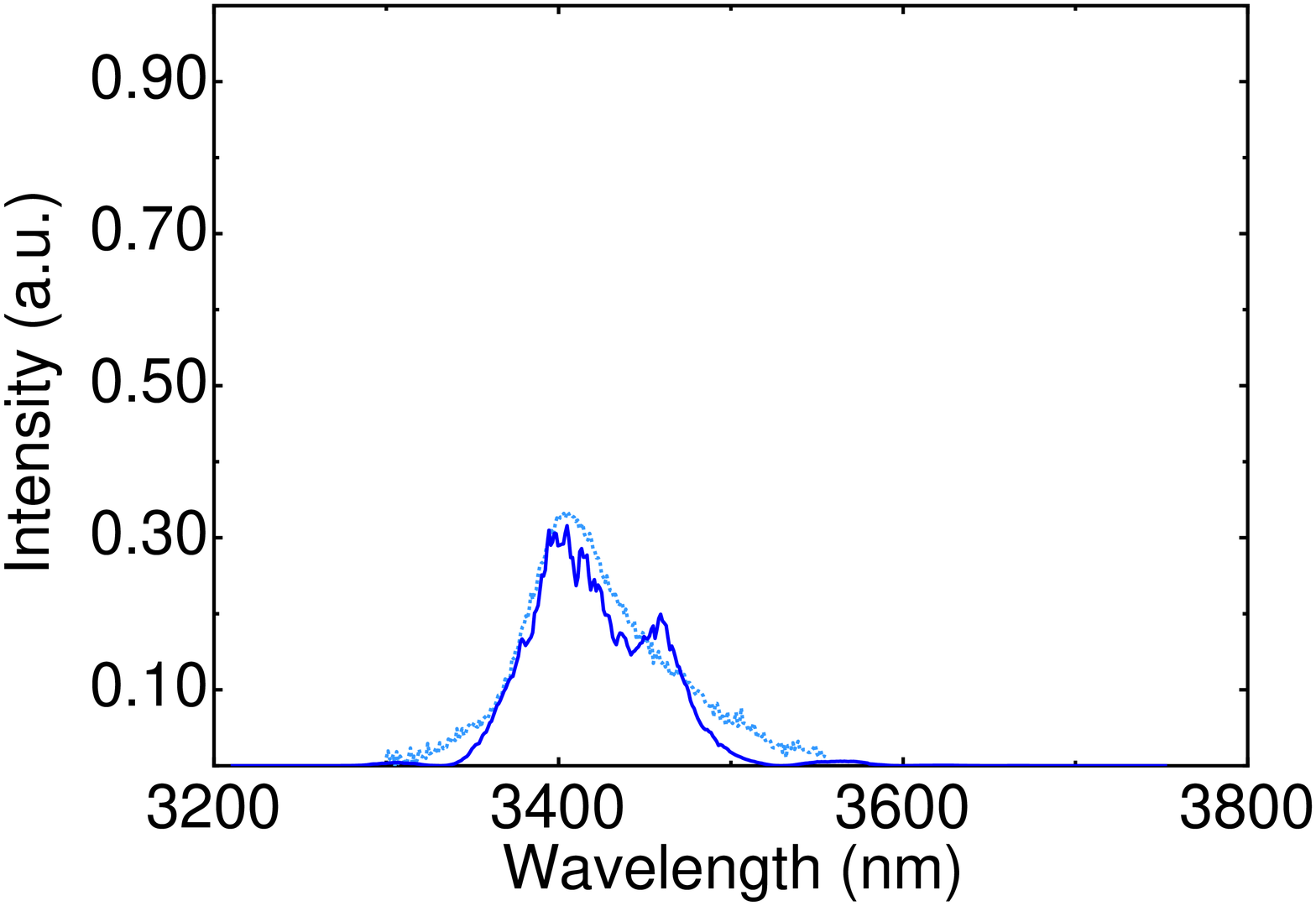}}
	 }
	 \caption{Subfigures (a)-(c) are calculated (dotted lines) and measured (solid lines) OPO idler spectra for different positions in the beam. (a), (b), and (c) correspond to the same locations as marked in Fig.~\ref{fig:OPOSpectrum}.
	 }
	 \label{fig:comparedSpectrum}
\end{figure}
 
The gain spectrum, when it is restricted to just one branch, agrees quite well with the experimentally observed idler spectrum, as can be seen in Fig.~\ref{fig:comparedSpectrum}, although the calculated idler spectrum is slightly broader. The calculated spectra in Fig.~\ref{fig:comparedSpectrum} are generated by summing the all the gain contributions for idler waves propagating at a particular angle. The idler angles in Fig.~\ref{fig:comparedSpectrum} correspond to 16.3 mrad, 11.1 mrad, and -13.7 mrad for subfigures (a), (b), and (c) respectively. The idler spectrum as a function of divergence angle is shown in more detail in Media~2. The differences in relative intensity is most likely due to changes in the experimental alignment as the beam was scanned across the pinhole. The difference between the calculated and observed spectra is believed to be due to fact that the model only takes into account phasematching considerations, pump divergence and the pump spectrum. It does not include the effects of spatial mode overlap, fast dynamic effects as described in refs.~\cite{Picozzi:2004ca} and \cite{Picozzi:2001hp}, or pump depletion, which will also modify the spectrum.


\section{Conclusion}
In conclusion, we have demonstrated a continuous wave OPO with, to our knowledge, the broadest output idler spectrum to date. A bandwidth of 230~nm (190~cm$^{-1}$) was achieved through choosing wavelengths and poling periods and focusing conditions to obtain a broad collinear and non-collinear phasematching. Our calculations indicate that with the right pumping conditions, an idler spectral bandwidth of up to 400~nm may be obtained with an ASE source that has a flatter emission spectrum. Furthermore, the non-collinear phasematching approach may be suited to provide broadened spectra in other idler wavelength ranges, where broadening in a collinear geometry is not possible, i.e., where $d\lambda_{s}/d\lambda_{p}\approx0$ is not fulfilled 

\section{Acknowledgements}
We gratefully acknowledge the useful discussions and preliminary work of Petra Gro\ss\ and Marvin Klein.
\end{document}